\documentclass[12pt,a4paper]{article}

\usepackage{fullpage}

\usepackage{graphicx}
\usepackage{amssymb,amsfonts,amsmath}
\usepackage{algorithm}
\usepackage{algcompatible}

\def\rd{{\rm d}}

\title{Quantifying Age and Model Uncertainties in Palaeoclimate Data and Dynamical Climate Models with a Joint Inferential Analysis}

\author{J. Carson$^{1}$\footnote{Correspondence: Jake.Carson@warwick.ac.uk} , M. Crucifix$^{2}$, S. P. Preston$^{3}$ and R. D. Wilkinson$^{4}$}

\date{}

\begin{document}

\maketitle



 $^{1}$Department of Statistics, University of Warwick, Coventry CV4 7AL, United Kingdom  \\

 $^{2}$Earth and Life Institute, Universite catholique de Louvain, 1348 Louvain-la-Neuve, Belgium  \\

 $^{3}$School of Mathematical Sciences, University of Nottingham, Nottingham NG7 2RD, United Kingdom \\

 $^{4}$School of Mathematics and Statistics, University of Sheffield, Sheffield S10 2TN, United Kingdom   \\


\begin{abstract}
The study of palaeoclimates relies on information sampled in natural archives such as deep sea cores. 
Scientific investigations often use such information in multi-stage analyses, typically with an age model being fitted to a core to convert depths into ages at stage one.
These age estimates are then used as inputs to develop, calibrate, or select climate models in a second stage of analysis. 
Here we show that such multi-stage approaches can lead to misleading conclusions,
and develop a joint inferential approach for climate reconstruction, model calibration, and age estimation.
As an illustration, we investigate the glacial-interglacial cycle, fitting both an age model and dynamical climate model to two benthic sediment cores spanning the past 780 kyr.
To show the danger of a multi-stage analysis we sample ages from the posterior distribution, then perform model selection conditional on the sampled age estimates, mimicking standard practice. 
Doing so repeatedly for different samples leads to model selection conclusions that are substantially different from each other, and from the joint inferential analysis. 
We conclude that multi-stage analyses are insufficient when dealing with uncertainty, and that to draw sound conclusions the full joint inferential analysis should be performed.
\end{abstract}












\maketitle

\section{Introduction}

Our understanding of palaeoclimates is based on data taken from climate archives that are proxy for climatic variables, such as temperature, as well as models of the climate that mathematically formulate hypotheses about long-term climate dynamics.
Sediment cores are commonly used climate archives, with measurements from different depths in the core relating to the climate at different points in time.
These data can be used to produce reconstructions of past climates by converting depths into ages through the use of an \textit{age model},
but doing so is a difficult process as the age-depth relationship is nonlinear, and there are typically only a few features present in a climate archive that can be used directly for dating.
Mathematical models are used to investigate climate dynamics, and are often calibrated using these reconstructions in a separate stage of analysis.
Care must be taken to avoid circular reasoning, as any inferences from the model calibration might otherwise result from assumptions artificially embedded in the age model \cite{Muller1997}.

There are numerous sources of uncertainty in each stage of the analysis;
for example uncertainties in the age and parameter estimates,  discrepancies between models and real-world system dynamics, and in how proxy measurements relate to climatic variables. 
Accurately quantifying these uncertainties, and in particular propagating uncertainties through the entire analysis, is essential if we are to trust in the conclusions from these investigations \cite{Blaauw2012}.
A multi-stage analysis offers no natural way to do so when strong dependencies exist between stages.
The aim of this article is to demonstrate that a single joint inferential analysis of the problem can and should be performed. 
This has been made possible by advances in computational Bayesian statistical methodology that allow us to simultaneously solve the probability calculus for all of the unknowns.
Using a joint inferential analysis avoids issues with circular reasoning, and ensures that uncertainties are  propagated correctly throughout the investigation. 

Our motivating example is the study of the glacial-interglacial cycle over the past 780 kyr.
Over this period the climate oscillated between cold periods in which glaciers extended, and warm periods in which the glaciers retreated \cite{Shackleton1984}.
This is clear in, for example, benthic cores of $\delta^{18}\mbox{O}$, which is a measurement of the ratio between $^{18}\mbox{O}$ and $^{16}\mbox{O}$ taken from calcite shells embedded in deep-sea sediment cores, and is primarily a function of global temperature and ice volume at the time the calcite shell was deposited \cite{Emiliani1955,Shackleton1967}.
The tasks we aim to perform are fitting an age model to the sediment cores (age estimation),
reconstructing components of the climate over time (climate reconstruction),
estimating the parameters of a climate model (model calibration),
and determining which models are best supported by the data (model selection).
This is in some sense the statistical holy grail for analysing this problem \cite{Tingley2012}, and has not been achieved before now (even for simple models) due to the computational complexities of such an approach.

Numerous climate reconstructions over this period have been obtained by averaging $\delta^{18}\mbox{O}$ measurements over multiple cores (known as \textit{stacking}), and then fitting an age model \cite{Imbrie1984,Lisiecki2005,Huybers2007}.
In line with Milankovitch theory \cite{Milankovitch1941} (translation in \cite{Milankovitch1998}), the age models have usually relied on astronomical tuning,  aligning features in the archives to variations in the Earth's orbit over time \cite{Imbrie1984,Lisiecki2005}.
Alternative approaches not relying on astronomical tuning have been developed with the aim of verifying the Milankovitch hypothesis \cite{Huybers2007,Huybers2004}.

Phenomenological models of the glacial-interglacial cycle are often characterized as either ordinary differential equations (ODEs) or stochastic differential equations (SDEs) that explicitly model a small number of climatic variables \cite{Crucifix2012,Crucifix2013}.
These are consistent with the underlying dynamics of the system, but are not analytically derived from the laws that govern the physical processes.
The models are typically astronomically forced, and so in a multi-stage analysis there is a clear risk of circular reasoning from calibrating such models using astronomically tuned age estimates. 
This danger has been demonstrated for model calibration and selection, which are extremely sensitive to the age estimates \cite{Feng2015,Carson2018}.
However, this sensitivity is apparent even when different sets of age estimates are consistent with the estimated age uncertainty in the data \cite{Carson2018}: two sets of age estimates that differ by an amount that is less than the error in the age estimates lead to conflicting conclusions about which models are more strongly supported by the data.
This shows that fixing the age estimates and ignoring the age uncertainty can severely bias the results of a multi-stage analysis, regardless of the choice of age model.
Due to strong mutual dependencies between the age estimates and the climate model, namely that the forcing in the climate model constrains the age estimates, and the age estimates influence the amount of forcing inferred in the model calibration, we must account for the uncertainty in these investigations by using a joint inferential analysis.

In this article we develop an approach for joint age, state, and parameter estimation, involving models of the climate and sediment accumulation, and proxy measurements taken from sediment cores.
The output is a sample of age estimates, reconstructions, and parameter estimates, that characterise the uncertainty in the inference. 
The approach is tested on synthetic data, and then applied to two cores from the Ocean Drilling Program, ODP677 \cite{Shackleton1990} and ODP846 \cite{Mix1995}, so that the age estimates from these cores can be compared with those of \cite{Lisiecki2005} and \cite{Huybers2007}.

An additional output and advantage of our approach is that it allows for the estimation of the model evidence, which can be used for model selection.
We investigate the impact of ignoring age uncertainty by sampling age estimates from the joint inferential analysis, and then keeping them fixed in a subsequent computational analysis in order to mimic a multi-stage analysis.
We demonstrate that conclusions differ greatly between different sampled age estimates, further motivating the need for a joint inferential analysis.

\section{Data and Models}

\begin{figure}[t]
\begin{center}
\centerline{\includegraphics[width=1\textwidth]{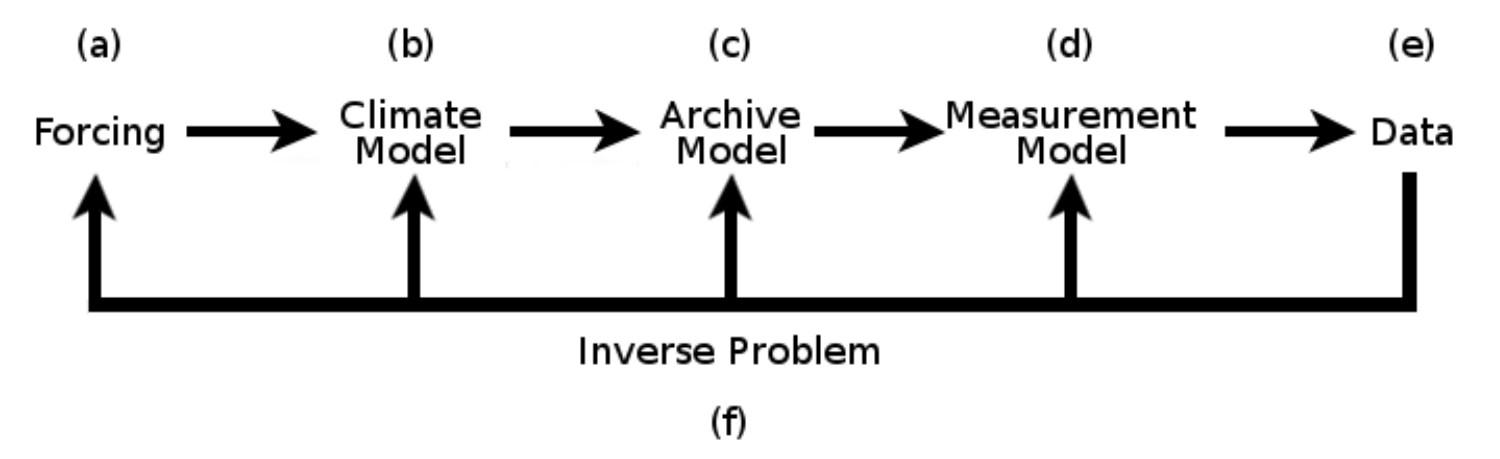}}
\caption{Illustration of the inverse problem. A set of models are used to describe the data generation process, and then the data are used to learn about these models.}
\label{Fig:MA}
\end{center}
\end{figure}

Our approach uses forward-models of each aspect of the system that results in the palaeoclimate data, and then learns all unknown quantities in a joint inferential analysis. 
The components we include are an astronomical forcing model (a) that drives a climate model (b); 
an archive model (c) that incorporates an age model relating core depths to ages, and a function linking proxy and climate variables;
and a measurement model (d) relating observations (e) to the true values.
This is summarized in Figure 1.
The models used here are relatively simple, but even for these  the inference is computationally challenging.
However, in principle each component could be replaced by a more sophisticated choice, and, notwithstanding computational challenges, the inference methodology described in this article could still be used.

\subsection{Forcing}

The prevailing theory is that the glacial--interglacial cycle is primarily driven by the seasonal and spatial variation of incoming solar radiation, termed ``insolation'', due to variations in the Earth's orbit around the Sun. 
The orbit is characterized by precession, obliquity, and eccentricity.
Precession refers to the angle, $\varpi$, made between the point of perihelion (the point of the orbit when the Earth is closest to the Sun) and the vernal point marking the spring equinox, and as such determines when in the seasonal cycle the Earth is closest to the Sun.
Obliquity is the angle between the equator and the orbital plane, and determines the insolation contrast between summer and winter.
Eccentricity measures how much the Earth's orbit deviates from  a perfect circle (indicated by zero eccentricity), and hence modulates the effect of precession.
It is often convenient to refer to {\it climatic precession}, $e\sin\varpi$, which combines the effects of eccentricity and precession in order to indicate the effect on the Northern Hemisphere summer insolation.
Climatic precession can be complemented with {\it coprecession}, $e\cos\varpi$, to effectively compute insolation at any time of year at any latitude \cite{Berger1978}.
In this article we compute these quantities using the algorithm in \cite{Berger1978}, which is suitably accurate over the past $\sim$ 1 Myr.

Phenomenological models are typically forced by an astronomical forcing function that summarizes the effect of the seasonal and spatial distribution of insolation \cite{Crucifix2013}. 
Here we use a forcing function of the form 
\begin{equation*}\label{eqn:forcing}
F(t; \boldsymbol{\gamma}) = \gamma_P \Pi_P (t) + \gamma_C \Pi_C (t) + \gamma_E E (t),
\end{equation*}
\noindent where $\Pi_P (t)$, $\Pi_C (t)$, and $E (t)$, are the normalized climatic precession, coprecession, and obliquity respectively. 
The parameter $\boldsymbol{\gamma}=(\gamma_P, \gamma_C, \gamma_E)^\top$ weights the linear combination.
Particular choices of $\boldsymbol{\gamma}$ correspond to several forcing functions used in the literature, for example caloric summer insolation at $60^{\circ}$ N as advocated by \cite{Milankovitch1941}, or daily mean insolation at $60^{\circ}$ N summer solstice as used in \cite{Imbrie1980}.
The model is unforced when  $\boldsymbol{\gamma}$ is set to zero.

\subsection{Climate model}

The astronomical forcing alone does not explain all of the features of the glacial-interglacial cycle, and so
the internal dynamics of the climate system must also be considered \cite{Imbrie1980,Raymo1997}.
The approach we follow is to model the Earth's climate as a dynamical system forced by the variation in the insolation \cite{Crucifix2013}.
The complexity of the climate model is necessarily limited by the computational cost of generating model simulations.
Many simple phenomenological models have been proposed, which typically comprise of a small number of differential equations representing hypothesized relationships between different aspects of the climate.
Here we take the CR14 model (a modified version of \cite{Crucifix2012}), which treats the climate model as a forced oscillator, i.e., the Earth's climate would fluctuate between hot and cold periods in the absence of forcing, but the oscillation is paced by the astronomical forcing.
In addition, we represent atmospheric variability as a stochastic process, resulting in the following SDE (suppressing dependence on time)
\begin{equation*}
\begin{aligned}
\rd X_{1} & = -\left( \beta_0 + \beta_1 X_{1} + \beta_2 \left( X_{1}^3 - X_{1} \right) + \delta X_{2} + \right. \\
& \hspace{120pt} \left.  F\left(\boldsymbol{\gamma}\right) \right) \rd t + \sigma_1 \rd W_{1} \\
\rd X_{2} & = \alpha \delta \left(X_{1} + X_{2} - \frac{X_{2}^3}{3} \right) \rd t + \sigma_2 \rd W_{2} ,
\end{aligned}
\end{equation*}
\noindent in which $X_{1}$ is taken to be ice volume, and $X_{2}$ is a non-physical variable acting to switch between glacial and interglacial states. 
The parameter $\beta_0$ scales the ice accumulation/ablation process depending on the sign, $\beta_1$ scales a linear feedback process, $\beta_2$ stabilizes the system when positive, $\delta$ is an inverse time scale, $\alpha$ controls the ratio of time scales between $X_{1}$ and $X_{2}$, and $\sigma_1$ and $\sigma_2$ scale the stochastic fluctuations.

\subsection{Archive model} 

There are no direct measurements of the Earth's climate or ice sheet extent over the time scale of interest. 
Information about the past state of the climate is stored in climate archives.
For example, the quantity of $\delta^{18}\mbox{O}$ in the ocean varies as a function of global temperature and sea ice extent.
This information can be extracted from sediment cores, for example, in order to study the climate history.
Our model for this process, termed the archive model, consists of two components: A model relating the proxy and climate variables, and an age model governing the age-depth relationship.
In general, larger values of $\delta^{18}\mbox{O}$ are indicative of a cold climate with a large amount of sea ice.
Hence for the proxy model we assume a simple linear relationship between ice volume ($X_{1}$), and $\delta^{18}\mbox{O}$ (denoted $Z$), i.e.,
\[ Z = D + C X_{1}, \]

The $\delta^{18}\mbox{O}$ datasets we use in this article are taken from benthic sediment cores.
Sediment accumulation is inherently stochastic, so as a starting point for the age model we model sediment accumulation by
\begin{equation}
\label{E:Sed}
\rd S = \mu_s \rd t + \sigma_s \rd W_s,
\end{equation}
\noindent where $S$ is the amount of sediment (in m), $W_s$ a standard Brownian motion, and the parameters $\mu_s$ and $\sigma_s$ represent the mean and variance of the sediment accumulation process.
Under this model, sediment can accumulate and dissipate, but when $\mu_s > 0$ the trend is for linear accumulation over time.

Equation \ref{E:Sed} can be inverted to give a model for the time, $T_m$, corresponding to some core depth $H_m$, where $1 \leq m \leq M$ identifies core slices.
Notationally we take larger values of $m$ to be nearer the top of the core ($H_{m} < H_{m-1}$), $T=0$ as the present, $S=0$ as the present sediment level, and $H=0$ as the top of the core.
When a core is sampled at depth $H_m$, the climate information recorded  corresponds to the most recent time at which $S = -H_m$. 
This gives a first passage time problem under the time reversal of Equation \ref{E:Sed}, the solution of which is an inverse Gaussian distribution, so that
\begin{equation}
\label{E:IG}
T_m - T_{m-1} \mid T_m \; \sim \; IG \left(\frac{H_{m-1} - H_m}{\mu_s}, \frac{\left(H_{m-1} - H_m \right)^2}{\sigma_s^2} \right).
\end{equation}
We can obtain $p \left( T_{m} \mid T_{m-1} \right)$ using Bayes theorem by noting that $p\left( T_{m}\right)$ follows Equation \ref{E:IG} conditioned on present values. 
Note that even though our sediment accumulation model is non-monotonic, age monotonically increases with core depth.

Following the sediment accumulation process the sediment is subject to post-deposition effects such as core compaction, which we can include by extending the model.
In order to model compaction, we utilize the linear porosity model of \cite{Huybers2007}, which should be suitable for the core depths of interest \cite{Bahr2001}.
This compaction adjustment introduces two additional parameters that need to be inferred, the gradient, $c$, and the intercept, $\phi_0$. 
The  model transforms depth measurements into a non-compacted equivalent, using 
\begin{equation*}
 \hat{H}_m = H_m + \frac{c}{1-\phi_0} H_m^2,
\end{equation*}
\noindent which can then be substituted into Equation \ref{E:IG}.

\subsection{Measurement model}

The final component in the forward modelling approach relates the output of the climate model to measurements taken from palaeoclimate archives.
For $\delta^{18}\mbox{O}$ we use Gaussian white noise measurement error,
\[ Y_m \sim \mathcal{N} ( Z_m, \sigma_y^2), \]
\noindent where $Y_m$ denotes the observation at depth $H_m$, $Z_m = Z(T_m)$, and $\sigma_y$ scales the amount of measurement error. 
In addition, there are often features in cores that have been independently dated, providing valuable information for age estimation.
Geomagnetic reversals, for instance, have been independently dated to high accuracy, and are frequently observable in benthic cores.
However, these are rare events, with the most recent being the Brunhes-Matuyama (BM) reversal, which occurred approximately 780 kyr ago.
Where the BM reversal can be observed within a core, we use 780 kyr as the age estimate of the associated $\delta^{18}\mbox{O}$ observation, and assume Gaussian error with a standard deviation of 2 kyr \cite{Singer1996}.

\subsection{Data}

We use the ODP677 \cite{Shackleton1990} and ODP846 \cite{Mix1995} benthic sediment cores.  
Both an astronomically tuned age model \cite{Lisiecki2005} and a non-astronomically tuned age model \cite{Huybers2007} have been fitted to each core, giving two different sets of age estimates for each core for comparison.
The BM reversal is identifiable in both ODP677 (at 30.4 m) and ODP846 (at 28.7 m), and so we take the measurements at these depths as the starting values.
The number of observations following the BM reversal are $M=363$ and $M=308$, for ODP677 and ODP846 respectively.

\subsection{Inverse problem}

We employ a Bayesian approach that simultaneously estimates ages, model parameters, climate states, and chooses between models by estimating Bayes factors. 
Formally, we target the posterior distribution
\begin{equation}
\label{E:Target}
p \left(\boldsymbol\theta, T_{1:M}, \boldsymbol{X}_{1:M} \mid Y_{1:M} \right) \propto p \left( Y_{1:M} \mid \boldsymbol\theta, T_{1:M}, \boldsymbol{X}_{1:M} \right) p \left( \boldsymbol{X}_{1:M} \mid \boldsymbol\theta, T_{1:M} \right) p \left( T_{1:M} \mid \boldsymbol\theta \right) p \left( \boldsymbol\theta \right),
\end{equation}
\noindent where $p \left( Y_{1:M} \mid \boldsymbol\theta, T_{1:M}, \boldsymbol{X}_{1:M} \right)$, $p \left( \boldsymbol{X}_{1:M} \mid \boldsymbol\theta, T_{1:M} \right)$, and $p \left( T_{1:M} \mid \boldsymbol\theta \right)$ are densities induced by the forward models described in (a)--(d), and $p \left( \boldsymbol\theta \right)$ is the user defined prior distribution. Where possible we base our prior distribution on physical grounds.
We constrain $\gamma_P$ and $\gamma_E$ to be positive in line with Milankovitch theory, which suggests that a positive northern hemisphere insolation anomaly in summer encourages a reduction in ice volume over time \cite{Berger2004}. 
Specifically, we choose  exponential prior distributions for these parameters, allowing for the system to be weakly forced.
The final astronomical forcing parameter, $\gamma_C$, influences the seasonal distribution of insolation.
Since we lack knowledge about whether more insolation in spring at the expense of autumn results in a positive or negative contribution to ice accumulation,
we choose a Gaussian prior distribution centred on zero (indicating summer solstice insolation).
The prior distributions  for the age model parameters, $\mu_s$ and $\sigma_s$, were selected by examining the sediment cores presented in \cite{Lisiecki2005,Huybers2007}, and the prior distributions  for the compaction adjustment parameters, $\phi_0$ and $c$, were chosen based on the porosity profiles presented in \cite{Huybers2004}.
The remaining parameters do not represent measurable quantities, making prior distribution specification  on physical grounds impossible.
Instead, prior distributions were selected based on trial simulations, so that undesirable behaviours such as non-oscillating regimes, having extremely short or long cycles, and numerical instabilities, were all discouraged.
The complete set of prior distributions for all three models is given in Table~\ref{Tab:Priors}.

\begin{table}[ht]
\begin{center}
\caption[List of parameters used to generate data for the simulation study, and the associated prior distributions used in the statistical analysis]{List of parameters used to generate data for the simulation study, and the associated prior distributions used in the statistical analysis.}
\label{Tab:Priors}
\begin{tabular}{ c  c  c }
Parameter & True Value & Prior Distribution \\*[1pt]
\hline
$\beta_0$ & 0.65 & $\mathcal{N}\left(0.4,0.3^2\right)$ \\*[1pt]
$\beta_1$ & 0.2 & $\mathcal{N}\left(0,0.4^2\right)$ \\*[1pt]
$\beta_2$ & 0.5 & $exp\left(1 \diagup 0.5 \right)$ \\*[1pt]
$\delta$ & 0.5 & $exp\left(1 \diagup 0.5 \right)$ \\*[1pt]
$\alpha$ & 11 & $\Gamma\left(10,2\right)$\\*[1pt]
$\gamma_P$ & 0.2 & $exp\left(1 \diagup 0.3 \right)$ \\*[1pt]
$\gamma_C$ & 0.1 & $\mathcal{N}\left(0,0.3^2\right)$ \\*[1pt]
$\gamma_E$ & 0.3 & $exp\left(1 \diagup 0.3 \right)$ \\*[1pt]
$\sigma_1$ & 0.2 & $exp\left(1 \diagup 0.3 \right)$ \\*[1pt]
$\sigma_2$ & 0.5 & $exp\left(1 \diagup 0.5 \right)$ \\*[1pt]
$\sigma_Y$ & 0.1 & $exp\left(1 \diagup 0.1 \right)$ \\*[1pt]
$D$ & 4.2 & $\mathcal{U}\left(3,5\right)$\\*[1pt]
$C$ & 0.8 & $\mathcal{U}\left(0.5,2\right)$\\*[1pt]
$\mu_s$ & $4.5 \times 10^{-5}$ & $\Gamma \left(180,1 \diagup 4 \times 10^6 \right)$\\*[1pt]
$\sigma_s$ & $2 \times 10^{-3}$ & $ exp(500)$\\*[1pt]
$\phi_0$ & $0.8$ & $\beta(45,15)$\\*[1pt]
$c$ & $3.5 \times 10^{-4}$ & $exp(4000)$\\*[1pt]
$X_1(t_1)$ & $-1$ & $\mathcal{U} \left(-1.5, 1.5\right)$\\*[1pt]
$X_2(t_1)$ & $-1.5$ & $\mathcal{U} \left(-2.5, 2.5\right)$\\*[1pt]
\hline
\end{tabular}
\end{center}
\end{table}



\section{Methods}

Since the posterior distribution has no analytical solution, we use a Monte Carlo approach that characterizes the posterior distribution with a large number of random samples, each one of which consists of a set of parameter values, climate reconstructions, and age estimates. 
Specifically we employ the sequential Monte Carlo squared (SMC$^2$) algorithm \cite{Chopin2013}, which was previously implemented in \cite{Carson2018} to test between competing phenomenological models of the glacial-interglacial cycle in the absence of any age uncertainty.
Here we extend the target distribution to also estimate the ages. 
SMC$^2$ is advantageous as it requires little user input in selecting tuning parameters, and so can be applied to multiple models and data sets with relative ease.


Sequential Monte Carlo (SMC) algorithms \cite{DelMoral2006} sample a target distribution, $p_M$, by propagating a weighted collection of `particles' through a series of intermediary distributions, $\{ p_m \}_{m=1}^M$.
The particles are initially sampled from an arbitrary tractable distribution, $p_1$, and the intermediary distributions are then chosen so as to gradually morph from $p_1$ to $p_M$.
The gradual transition between distributions allows for the implementation of an efficient sampling scheme in every iteration.
The reason for using SMC methods is that they provide an unbiased estimate of the normalizing constant for each distribution that can then be used in parameter estimation or model selection.

\begin{algorithm}[p]
{\footnotesize
\caption[Particle filter algorithm for joint state and observation age estimation]{Particle filter targeting $p \left(T_{1:M}, \boldsymbol{X}_{1:M} \mid Y_{1:M}, \boldsymbol{\theta} \right)$.}
\label{A:PFT}


\begin{algorithmic}

\FOR{$k=1,...,N_X$}
\STATE{Sample $T_1^{(k)} \sim b_1 \left(T_1 \mid Y_1, \boldsymbol{\theta} \right)$.}
\STATE{Sample $\boldsymbol{X}_1^{\left(k \right)} \sim r_1\left(\boldsymbol{X}_1 \mid T_1^{(k)}, Y_1, \boldsymbol{\theta} \right)$.}
\STATE{Set the importance weight
\[ \omega_1^{(k)} = \frac{p \left(T_1^{(k)} \mid \boldsymbol{\theta} \right) p \left(\boldsymbol{X}_1^{\left(k \right)} \mid T_1^{(k)}, \boldsymbol{\theta} \right) p \left(Y_1 \mid T_1^{(k)}, \boldsymbol{X}_1^{\left(k \right)}, \boldsymbol{\theta} \right)}{b_1 \left(T_1^{(k)} \mid Y_1, \boldsymbol{\theta} \right) r_1\left(\boldsymbol{X}_1^{\left(k \right)} \mid T_1^{(k)}, Y_1,  \boldsymbol{\theta} \right)}. \]}
\ENDFOR
\STATE{Normalize the weights. For $k=1,...,N_X$
\[ \Omega_1^{(k)} = \frac{\omega_1^{(k)}}{\sum_{i=1}^{N_X} \omega_1^{(i)}}. \]}
\FOR{$m=2,...,M$}
\FOR{$k=1,...,N_X$}
\STATE{Sample ancestor particle index $a_{m-1}^{(k)}$ with replacement from $1:N_x$ according to weights $\Omega_{m-1}^{(1:N_X)}$.}
\STATE{Sample $T_m^{(k)} \sim b_m \left(T_m \mid  T_{m-1}^{\left(a_{m-1}^{(k)} \right)}, Y_{m}, \boldsymbol{\theta} \right)$.}
\STATE{Sample $\boldsymbol{X}_m^{\left(k \right)} \sim r_m\left(\boldsymbol{X}_m \mid T_{m-1}^{\left(a_{m-1}^{(k)} \right)}, T_m^{(k)}, \boldsymbol{X}_{m-1}^{\left(a_{m-1}^{(k)} \right)} , Y_m,  \boldsymbol{\theta} \right)$.}
\STATE{Extend the particle trajectory \[ \left\lbrace T_{1:m}^{(k)},\boldsymbol{X}_{1:m}^{(k)} \right\rbrace = \left\lbrace \left(T_{1:m-1}^{(a_{m-1}^{(k)})}, T_m^{(k)} \right), \left(\boldsymbol{X}_{1:m-1}^{(a_{m-1}^{(k)})}, \boldsymbol{X}_m^{\left(k \right)} \right) \right\rbrace . \]}
 \STATE{Set the importance weight
\begin{multline*}
\omega_m^{(k)} = p \left( Y_m \mid \boldsymbol{X}_m^{\left(k \right)}, \boldsymbol{\theta} \right) \frac{p\left(T_m^{(k)} \mid T_{m-1}^{\left(a_{m-1}^{(k)} \right)}, \boldsymbol{\theta} \right)}{b_m \left(T_m^{(k)} \mid  T_{m-1}^{\left(a_{m-1}^{(k)} \right)}, Y_{m}, \boldsymbol{\theta} \right) } \times \\
\frac{p \left(\boldsymbol{X}_m^{\left(k \right)} \mid \boldsymbol{X}_{m-1}^{\left(a_{m-1}^{(k)} \right)}, T_m^{(k)}, T_{m-1}^{\left(a_{m-1}^{(k)}\right)}, \boldsymbol{\theta} \right)}{r_m\left(\boldsymbol{X}^{(k)}_m \mid T_{m-1}^{\left(a_{m-1}^{(k)} \right)}, T_m^{(k)}, \boldsymbol{X}_{m-1}^{\left(a_{m-1}^{(k)} \right)} , Y_m,  \boldsymbol{\theta} \right)}.
\end{multline*}
}
\ENDFOR
\STATE{Normalize the weights. For $k=1,...,N_X$
\[ \Omega_m^{(k)} = \frac{\omega_m^{(k)}}{\sum_{i=1}^{N_X} \omega_m^{(i)}}. \]}
\ENDFOR
\end{algorithmic}
}
\end{algorithm}

For state space models it is common to initialize by setting $p_1$ to be  the prior distribution, and to then add a single data point for every intermediary distribution.
A well known example is the particle filter (PF) \cite{Gordon1993}, which uses the sequence of filtering distributions $p_m(\boldsymbol{X}_{1:m}) := p(\boldsymbol{X}_{1:m} \mid Y_{1:m}, \boldsymbol{\theta})$.
Particles are sampled and propagated via importance sampling, so that at initialization a sample of $N_x$ particles are sampled from some proposal density $r_1(\boldsymbol{X}_1 \mid Y_1, \boldsymbol{\theta})$, and given importance weight $p (\boldsymbol{X}_1, Y_1 \mid \boldsymbol{\theta}) \diagup r_1(\boldsymbol{X}_1 \mid Y_1, \boldsymbol{\theta})$.
In subsequent iterations the particles are resampled using a multinomial scheme, propagated via some proposal distribution $r_m(\boldsymbol{X}_m \mid \boldsymbol{X}_{1:m-1}, Y_{1:m}, \boldsymbol{\theta})$, and reweighted so that the particles are a weighted sample of the posterior $p(\boldsymbol{X}_{1:m} \mid Y_{1:m}, \boldsymbol{\theta})$.

Here we extend the target distribution to also sample the ages, i.e., we use the sequence of intermediary distributions $p_m(\boldsymbol{X}_{1:m}, T_{1:m}) = p(\boldsymbol{X}_{1:m}, T_{1:m} \mid Y_{1:m}, \boldsymbol{\theta})$.
This in turn requires us to extend the proposal distribution.
Our approach is to use a two-step proposal, proposing $T_m$ from $b_m \left(T_m \mid  T_{m-1}, Y_{m}, \boldsymbol{\theta} \right)$, followed by $\boldsymbol{X}_m$ from $r_m\left(\boldsymbol{X}_m \mid T_{m-1}, T_m, \boldsymbol{X}_{m-1}, Y_m, \boldsymbol{\theta} \right)$.
The full algorithm is described in Algorithm~\ref{A:PFT}.

Using the PF, unbiased estimates of the normalizing constants  $p(Y_{1:m} \mid \boldsymbol{\theta})$ (termed the likelihood) can be obtained in each iteration by noting that this normalizing constant can be decomposed as
\begin{equation*}\label{eqn:Ident3}
p (Y_{1:m} \mid \boldsymbol{\theta}) = p(Y_1 \mid \boldsymbol{\theta}) \prod_{j=2}^m p (Y_j \mid Y_{1:j-1}, \boldsymbol{\theta}).
\end{equation*}
\noindent Unbiased estimates of each of the components can be obtained by averaging the unnormalised weights in the particle filter in each iteration, i.e., 
\begin{equation*}
\hat{p} (Y_m \mid Y_{1:m-1}, \boldsymbol{\theta}) = \frac{1}{N_x} \sum_{k=1}^{N_x} \omega_m^{(k)}. 
\end{equation*}
Taking the product of these unbiased estimates in turn gives an unbiased estimate, $\hat{p}(Y_{1:m} \mid \boldsymbol{\theta})$, of  $p(Y_{1:m} \mid \boldsymbol{\theta})$ \cite{DelMoral2004}.

At the conclusion of the PF (once all of the data have been assimilated), we have an unbiased estimate of the likelihood, $\hat{p}(Y_{1:M} \mid \boldsymbol{\theta})$.
The unbiased likelihood estimates can then be embedded within another Monte Carlo algorithm in order to perform parameter estimation (sampling from $p(\boldsymbol{\theta} \mid Y_{1:M})$).
These are termed pseudo-marginal algorithms, and are constructed in such a way as to target the correct posterior distribution, despite using the approximate likelihood $\hat{p}(Y_{1:M} \mid \boldsymbol{\theta})$ in the Monte Carlo scheme \cite{Andrieu2009}.
Recent examples include PMCMC \cite{Andrieu2010}, which embeds the PF within an MCMC algorithm, and SMC$^2$ \cite{Chopin2013}, which embeds the PF within a second SMC algorithm.
As we are interested in obtaining estimates of the normalizing constant $p(Y_{1:M})$, we focus on the SMC$^2$ algorithm.

The SMC$^2$ algorithm \cite{Chopin2013} embeds the particle filter within another SMC algorithm in order to target the sequence of posterior distributions
\begin{equation*}
p_0 = p(\boldsymbol{\theta}), \hspace{1cm} p_m = p(\boldsymbol{\theta}, \boldsymbol{X}_{1:m} \mid Y_{1:m}),
\end{equation*}
for $m=1, \ldots, M$. 
The first step samples $N_\theta$ parameter particles, $\{\boldsymbol{\theta}^{(n)}\}_{n=1}^{N_\theta}$, from the prior distribution, and attaches a PF of $N_x$ state particles to each parameter particle.
As observations are assimilated, the attached PFs return unbiased estimates of the likelihoods, $p(Y_{1:m} \mid \boldsymbol{\theta}^{(n)})$, which are used to weight the parameter particles. 
The weights must be carefully monitored in each iteration as a small number of particles tend to accumulate most of the weight, dominating the particle approximation.
This is referred to as particle degeneracy, and is often monitored by tracking the effective sample size (ESS), defined as
\[\mbox{ESS } = \left( \sum_{i=1}^{N_\theta}  \left( W_m^{(i)} \right)^2 \right)^{-1}, \] 
\noindent  where $\left\lbrace W_m^{(i)} \right\rbrace_{i=1}^{N_\theta}$ are the normalized weights in population $m$.
The particles can be resampled when the ESS falls below some threshold (usually $N_\theta / 2$), so that low-weight particles are discarded.
Doing so equalizes the weights between particles, but leads to few unique particles in the parameter space.
In SMC$^2$ particle diversity can be improved after the resampling step by running a PMCMC algorithm that leaves $p (\boldsymbol{\theta}, \boldsymbol{X}_{1:m} \mid Y_{1:m})$ invariant \cite{Andrieu2010},
i.e. new values are proposed for each parameter particle, and a PF performed up to time $m$ conditional on the proposed parameters.
The new parameter values are then accepted or rejected according to the ratio of the posterior density estimates. 
The full algorithm, extended to include $T_{1:M}$ in the target, is given in Algorithm~\ref{A:SMC2M}.

\begin{algorithm}[p]
{\footnotesize
\caption{SMC$^2$ algorithm targeting $p \left( \boldsymbol{\theta}, \boldsymbol{X}_{1:M}, T_{1:M} \mid Y_{1:M} \right)$.}
\label{A:SMC2M}

\begin{algorithmic}

\FOR{$n=1,...,N_{\theta}$}
\STATE{Sample $\boldsymbol{\theta}^{(n)}$ from the prior distribution, $p \left(\boldsymbol{\theta}\right)$.}
\STATE{Set the importance weight \[ W^{(n)}_0 = \frac{1}{N_\theta}. \]}
\ENDFOR

\FOR{$m=1,...,M$}
\IF{ESS$<\frac{N_{\theta}}{2}$}
\FOR{$n=1,...,N_{\theta}$}
\STATE{Sample $\boldsymbol{\theta}^{*(n)}$, $T_{1:m-1}^{*(1:N_X,n)}$ and $\boldsymbol{X}_{1:m-1}^{*(1:N_X,n)}$ from $\boldsymbol{\theta}^{(1:N_{\theta})}$, $T_{1:m-1}^{(1:N_X,1:N_{\theta})}$ and $\boldsymbol{X}_{1:m-1}^{(1:N_X,1:N_{\theta})}$, according to weights $W_{m-1}^{(1:N_{\theta})}$.}
\STATE{Sample $\boldsymbol{\theta}^{**(n)}$, $T_{1:m-1}^{**(1:N_X,n)}$ and $\boldsymbol{X}_{1:m-1}^{**(1:N_X,n)}$ from a PMCMC algorithm targeting  $p \left(\boldsymbol{\theta}, \boldsymbol{X}_{1:m-1}, T_{1:m-1} \mid Y_{1:m-1} \right)$ initialised with $\boldsymbol{\theta}^{*(n)}$, $T_{1:m-1}^{*(1:N_X,n)}$ and $\boldsymbol{X}_{1:m-1}^{*(1:N_X,n)}$.}
\ENDFOR 
\STATE{Set $\boldsymbol{\theta}^{(1:N_\theta)} = \boldsymbol{\theta}^{**(1:N_\theta)}$, $T_{1:m-1}^{(1:N_X,1:N_\theta)} = T_{1:m-1}^{**(1:N_X,1:N_\theta)}$ and $\boldsymbol{X}_{1:m-1}^{(1:N_X,1:N_\theta)} = \boldsymbol{X}_{1:m-1}^{**(1:N_X,1:N_\theta)}$.}
\STATE{Set the importance weights 
\[W_{m-1}^{(n)}=\frac{1}{N_{\theta}} \mbox{ for } n=1,...,n_\theta.\]}
\ENDIF 
\FOR{$n=1,...,N_{\theta}$}
\STATE{Sample $T_{1:m}^{(1:N_X,n)}$, $\boldsymbol{X}_{1:m}^{(1:N_X,n)}$ by performing iteration $m$ of the particle filter, and record estimates of $\hat{p} \left( Y_m \mid Y_{1:m-1}, \boldsymbol{\theta}^{(n)} \right)$ and $\hat{p} \left( Y_{1:m} \mid \boldsymbol{\theta}^{(n)} \right)$.}
\STATE{Set the importance weights
\[ w^{(n)}_m = W^{(n)}_{m-1} \hat{p} \left( Y_m \mid Y_{1:m-1}, \boldsymbol{\theta}^{(n)} \right). \]}
\ENDFOR
\STATE{Evaluate \[ \hat{p} \left(Y_m \mid Y_{1:m-1} \right) = \sum_{i=1}^{N_\theta} w_m^{(i)}. \]}
\STATE{Normalise the weights  \[ W^{(n)}_m = \frac{w^{(n)}_m}{\sum_{i=1}^{N_\theta} w^{(i)}_m}\; \mbox{ for }\; n=1,...,N_\theta. \]}
\ENDFOR

\end{algorithmic}
}
\end{algorithm}

SMC$^2$ also provides an estimate to the normalizing constant to Equation \ref{E:Target}, termed the model evidence.
As with the PF, SMC$^2$ makes use of the decomposition
\begin{equation*}
p (Y_{1:M} ) = p(Y_1) \prod_{m=2}^M p (Y_m \mid Y_{1:m-1}).
\end{equation*}
\noindent The components are then estimated by averaging the unnormalised weights in each iteration, i.e.,
\[ \hat{p}(Y_m \mid Y_{1:m-1}) = \sum_{n=1}^{N_\theta} W^{(n)}_{m-1} \hat{p} (Y_m \mid Y_{1:m-1}, \boldsymbol{\theta}^{(n)}), \]
\noindent and an estimate of the model evidence is then obtained by substituting $p(Y_m \mid Y_{1:m-1})$ with $\hat{p}(Y_m \mid Y_{1:m-1})$.
The ratio of model evidence terms between two models, i.e. $\mbox{$p(Y_{1:M} \mid \mathcal{M}_1) / p(Y_{1:M} \mid \mathcal{M}_2)$}$ where $\mathcal{M}_1$ and $\mathcal{M}_2$ are model identifiers, is the Bayes factor of model $\mathcal{M}_1$ over $\mathcal{M}_2$, and indicates the relative explanatory power between the two models \cite{Jeffreys1939}. 
Bayes factors are a commonly used tool for performing model selection in a Bayesian framework. Standard interpretations of the Bayes factor are described in \cite{Kass1995}. 
The Bayes factors provide a principled way to undertake model selection, such as comparing two different phenomenological models, or different astronomical forcings.

\subsection{Implementation details}
\label{SS:ID}

There are several user-defined choices to make when implementing the algorithm.
Firstly there are the proposal distributions in both the PF and the SMC algorithm in which it is contained.
An advantage of SMC approaches is that we have a collection of particles in each iteration that permits some automation of the proposals.
For example, for the model parameters in the PMCMC component we use independent Gaussian proposals with the sample mean and covariance.
In the PF, in the first iteration we initialize with the proposals $b_1\left( T_1 \mid Y_1, \boldsymbol{\theta} \right) \sim \mathcal{N} (780, 2)$ (kyr ago), which is the observation error distribution of the Brunhes-Matuyama (BM) geomagnetic reversal, and $r_1\left( \boldsymbol{X}_1 \mid T_1, Y_1, \boldsymbol{\theta} \right) = p \left( \boldsymbol{X}_1 \right)$, which is the prior distribution.
Within the PMCMC steps, whenever the PF is reinitialized we replace these proposals with independent Gaussian proposals using the sample mean and covariance.
As the PF progresses, for the ages we use the proposal $b_m \left(T_m \mid  T_{m-1}, Y_{m}, \boldsymbol{\theta} \right) \propto p \left(T_{m-1} \mid T_m  \right)$, which is a reasonable approximation to simulating from the age model.
Developing proposal distributions for the state variables is more difficult as the transition densities 
$p (\boldsymbol{X}_m \mid T_{m-1}, T_m, \boldsymbol{X}_{m-1}, \boldsymbol{\theta})$ are not available in closed form for the models of interest.
One option is to choose the particle proposal distributions so that the transition density cancels from the importance weights, which can be achieved by simulating each $\boldsymbol{X}_m$ from the climate model, i.e., by setting $r_m= p(\boldsymbol{X}_m \mid T_{m-1}, T_m, \boldsymbol{X}_{m-1}, \boldsymbol{\theta})$.
However, this will typically lead to too many proposals being far from the observations, leading to particle degeneracy.
Instead we use the proposals developed in \cite{Carson2018} that guide proposals towards the next observation, giving more equal weights.

Secondly, we need to decide on a resampling scheme. Here we use multinomial resampling, which is the most commonly used resampling scheme, but alternatives such as stratified resampling usually give improvements in sample variance \cite{Liu1998,Douc2005}. 

Thirdly, we need to decide on the number of particles, $N_\theta$ and $N_x$, and the chain length for the PMCMC rejuvenation steps.
These choices are typically dictated by the available computational resources.
We use $N_x = N_\theta = 1000$, and a chain length of 10, which seems to maintain high particle diversity.

Finally, we check whether the algorithm has converged by ensuring that the results are consistent between independent runs.

\begin{figure}[t]
\centering
\includegraphics[width=0.9\textwidth]{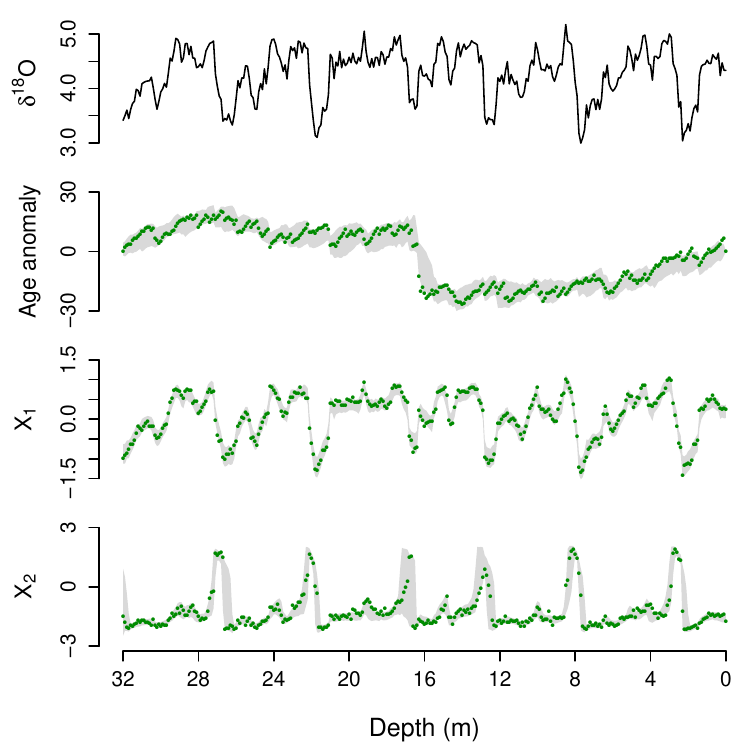}
\caption{Simulated data (black line) and estimated 95\% HDRs (grey) for the ages and states from the joint inferential analysis. Age estimates are shown with the linear trend removed. Points show true simulated values.}
\label{Fig:SS_Results}
\end{figure}

\begin{figure}[p]
\centering
\includegraphics[width=0.9\textwidth]{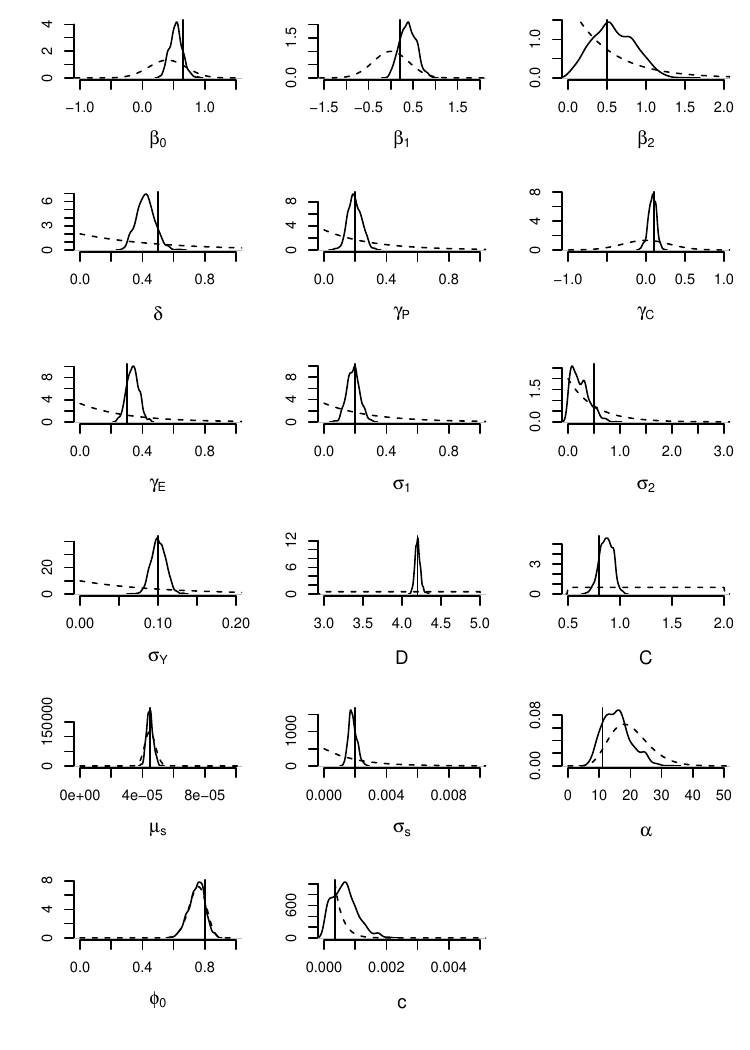}
\caption{Estimated marginal posterior distributions for the simulation study using the joint inferential analysis. Vertical lines indicate the values used to generate the data, and the dashed lines show the prior distributions.}
\label{Fig:SS_Results2}
\end{figure}

\begin{figure}[p]
\centering
\includegraphics[width=0.9\textwidth]{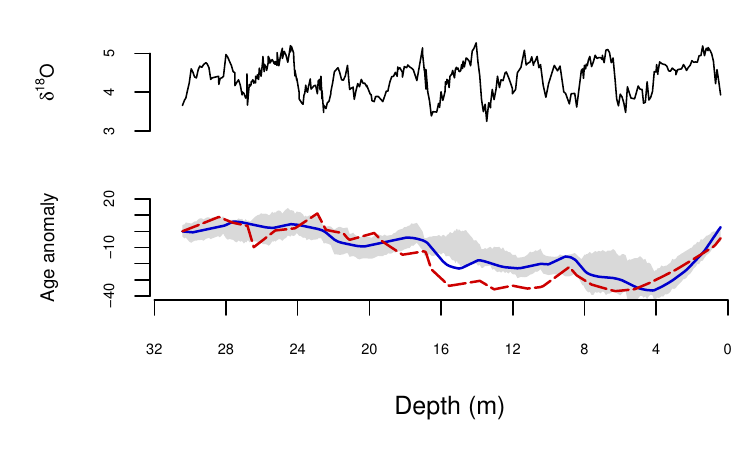}\\
\includegraphics[width=0.9\textwidth]{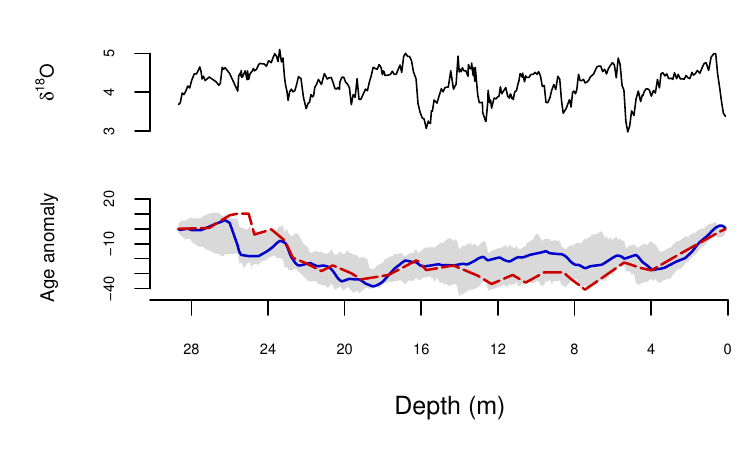}
\caption{Observed $\delta^{18}\mbox{O}$ values and age estimates from the joint inferential analysis for ODP677 (top), and ODP846 (bottom). 95\% HDRs for the observation ages are shown in grey. Age estimates from the LR04 benthic stack are shows as a solid line (blue in online version), and age estimates from the H07 stack are shown as a dashed line (red in online version).}
\label{Fig:ODP_Ages}
\end{figure}

\begin{figure}[p]
\centering
\includegraphics[width=0.9\linewidth]{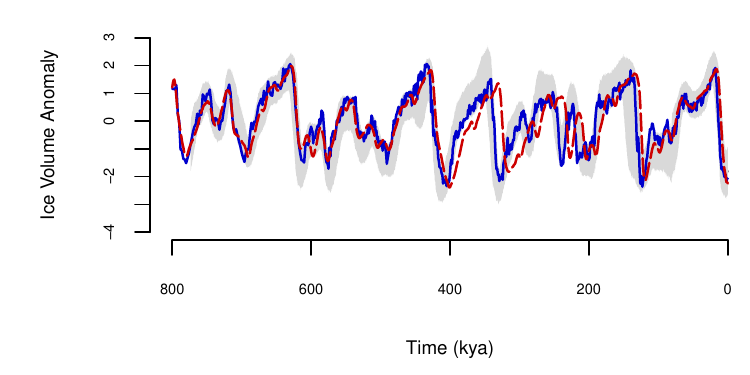}\\
\includegraphics[width=0.9\linewidth]{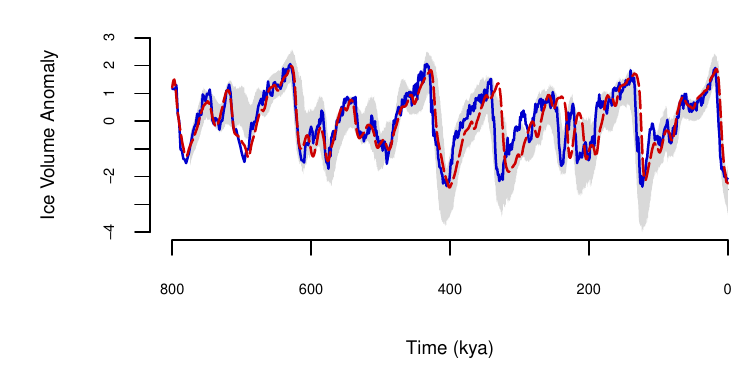}
\caption{95\% HDRs for the normalised ice volume over time from the joint inferential analysis for ODP677 (top) and ODP846 (bottom), shown in grey. The normalised LR04 stack is shown as a solid line (blue in online version), and the normalised H07 stack is shown as a dashed line (red in online version).}
\label{Fig:ODP_Recon}
\end{figure}

\section{Results}

\subsection{Joint inferential analysis reliably infers unobserved components from synthetic data}

We begin with a simulation study to demonstrate the ability of our joint inferential analysis for age estimation, state estimation, parameter estimation, and model selection.
We simulate a set of observations from  the forward models, and then attempt to recover the true parameters, states, and ages.
We also estimate the Bayes factors between the forced and unforced model to ensure that we can determine the importance of the astronomical forcing.
Specifically, observation times were drawn from an imagined core of length 32m sampled at 0.1 m intervals, giving $M=321$ observations. 
The first observation is a noisy measurement of the true age, where the noise is sampled from a Gaussian distribution with mean zero and a standard deviation of 2 kyr, as if we had observed the BM reversal.

Figure~\ref{Fig:SS_Results} shows the simulated observations, and  compares them with the 95\% highest density regions (HDRs) for the estimated ages and states, as well as the marginal posterior distributions of the parameters and their true values. 
The linear trend has been removed from the age versus depth plot so that the variation is more clearly visible. 
Note that the age-depth relationship is not a linear relationship (which would be a horizontal line), and in
 particular, there is a large period of time in which little sediment is deposited in the middle of the record. 
Despite this hiatus, the true ages are in regions of high posterior probability density throughout the dataset, showing that we are able to recover the ages.
Likewise, the majority of the true values for both the observable and unobservable state variables lie within the 95\% HDRs throughout the core. 
As would be expected, the posterior variance for the unobservable state is larger than for the observable state, particularly when the system switches between glacial and interglacial periods.
Finally, the true parameter values lie in regions of high posterior probability density.

We repeat the analysis for these data but now using an unforced version of the CR14 model. 
We focus on the Bayes factor to determine whether we can infer the importance of the astronomical forcing.
The Bayes factor is approximately $10^9$ in favour of the forced model, suggesting that even with the age uncertainty, the data strongly support the forced model \cite{Carson2018,Kass1995}. In other words, even though we have a relatively small number of noisy measurements from a single core, we are able to correctly infer the importance of the astronomical forcing in the climate record, even after accounting for the uncertainty that arises from estimating 17 model parameters, the age-depth relationship, and the climate states.
To ensure that this is a reasonable result, we can perform the model selection experiment on simulated data that has been generated with the unforced model.
In this case, we find that the Bayes factor is $10^2$ in favour of the unforced model, demonstrating that we are inferring the forcing parameters, and not simply assuming that the forcing plays a crucial role.

\subsection{Reconstructions from ODP677 and ODP846 are consistent with LR04 but not H07}

We now analyse ODP677 and ODP846, which are shown in Figure~\ref{Fig:ODP_Ages} with the estimated sequence of 95\% HDRs for the ages.
We include the age estimates from the LR04 \cite{Lisiecki2005} and H07 \cite{Huybers2007} stacks for comparison.
The age uncertainties are larger than in the simulation study, likely as a result of model discrepancies. 
Between the two cores, the age uncertainties are typically larger in ODP846 than ODP677, with the mean standard deviation of the age estimates being 6.5 kyr in ODP846 and 3.5 kyr in ODP677.
Both are smaller than previous uncertainty estimates, which were up to 11 kyr \cite{Lisiecki2005,Huybers2004,Huybers2007}.    
Additionally, the most uncertain estimates are not necessarily at the mid-point between age control points (such as the present, or geomagnetic reversals), which has previously been assumed \cite{Huybers2004,Huybers2007}. 
Rather, the age control points only seem to constrain the ages within a few meters of the core. 
Our age estimates for ODP677 are consistent with the LR04 age estimates, which lie in credible intervals throughout the sediment core. 
On the other hand, the H07 estimates deviate greatly from our estimates between 11 m and 16 m. 
Our age estimates for ODP846 are consistent with both LR04 and H07, primarily due to the larger variance in the age estimates. 
However, the LR04 estimates are notably closer to the posterior mean.  
For both datasets it can be seen that using a linear age-depth relationship will lead to poor estimates of observation times. 

The sequence of 95\% HDRs of the normalized ice volume vs time are shown in Figure~\ref{Fig:ODP_Recon}. 
We include the LR04 and H07 stacks for comparison.
It is reassuring that our ice-volume reconstructions from ODP677 and ODP846 are remarkably similar despite being obtained independently.
The similarities with LR04 are again very striking, whereas H07 is out of agreement between 200 and 400 kyr ago.
The likely reason is that the age estimates in the H07 stack are purely depth derived, and since this period is distant from the age-control points provided by the BM reversal and the present (core-top), the H07 reconstruction has low accuracy here.

Our approach has the advantage of using information on climate forcing, while preserving the possibility to test the alternative hypothesis that the astronomical forcing has no influence. 
Repeating the experiment using the unforced CR14 model yields Bayes factors in favour of the forced model against the unforced model of approximately $10^5$ in ODP677, and $1$ in ODP846. 
In other words, the forced model is strongly supported by ODP677, while evidence about astronomical forcing from ODP846 is weaker, but not contradictory. 

\subsection{Joint inferential analysis, as opposed to a multi-stage analysis, is essential}

In \cite{Carson2018} it was demonstrated that model selection experiments are sensitive to the choice of age model by showing that two sets of age estimates consistent with the reported uncertainty lead to different conclusions, and it was thus argued that the age uncertainty must be incorporated into any analysis.
In order to further illustrate this statement we propose the following experiment.
In performing a joint inferential analysis we have randomly generated realizations of the ages of each sediment core.
By sampling from these realizations we can obtain plausible age estimates that differ from each other consistently with the age uncertainty.
We can then repeat the analysis with the fixed age estimates in order to test whether our conclusions significantly differ when the age uncertainty is ignored.
Taking four realizations of the ages for ODP677, and then obtaining the Bayes factor in favour of the forced over the unforced model, yields Bayes factors of approximately  $10^4$, $10^8$, $10^9$, and $10^{15}$. 
Likewise for ODP846 we obtain Bayes factors of $10^{-2}$, $10^1$, $10^5$, and $10^8$; values less than one are evidence in favour of the unforced model.
In ODP677 we always favour the forced model, but with varying degrees of confidence.
In ODP846 our conclusions can change drastically, likely because the age uncertainty is greater.
In either case it is demonstrated that ignoring the age uncertainty can significantly alter the results of such analyses.
Due to strong couplings between the different components of the system, accurately characterizing and propagating uncertainties requires a joint inferential analysis.

\section{Conclusion}

We have investigated an approach to calibrating dynamical climate models and testing between competing hypotheses via model selection, whilst jointly fitting an age model to a sediment core.
Performing a joint inferential analysis in this manner is highly challenging, requiring state of the art statistical methods and intensive computation (the analyses presented here each took six days on a standard desktop computer).
Nevertheless there are notable advantages in undertaking a joint inferential analysis over splitting the analysis between multiple stages.
Firstly, the joint inferential analysis both estimates climate dynamics and forcings, and uses this information to constrain the ages, without the risk of circular reasoning.
Secondly, we are able to characterize a range of uncertainties, and push these uncertainties forward into our conclusions, making those conclusions more reliable.

There are several ways in which the approach presented here could potentially be extended.
Firstly, in each experiment we only utilize data from a single sediment core, and a natural extension is to combine observations from multiple cores.
However, this is a non-trivial extension; 
with multiple cores the order in time of the observations is unknown, and so developing effective proposal distributions in a Monte Carlo approach is significantly more difficult.
Secondly, numerous sources of uncertainty are not accounted for here.
Examples include identification error for the BM reversal, and bioturbation in the sediment core.
Such sources of uncertainty could be incorporated into the analysis by extending the models.
Finally, the models considered here are all relatively simple, and could be replaced with more complex models.
A primary obstacle in these extensions is the computational cost involved, which is already high when using simple models.
Fortunately, sequential Monte Carlo methods such as those used here are amenable to parallelization, giving the possibility of dramatically improving computation times.

Our focus for model selection was testing between a forced and an unforced model for the glacial--interglacial cycle, but the joint inferential approach is relevant to a much wider range of investigations.
Dynamical models can be used to investigate, for example, changes in oscillation regimes such as the mid-Pleistocene transition, abrupt changes during glacial periods such as Dansgaard--Oeschger events, and relationships between different climate variables such as ice volume and $\mbox{CO}_2$.
In each of these cases the calibration of the dynamical model will be sensitive to the inferred ages of the observations.
Testing between different hypotheses therefore requires effective joint quantification of age and model uncertainties, which can only be achieved by performing a joint inferential analysis.

\subsubsection*{Data Accessibility}
Code and data supporting this manuscript are available from Zenodo \hfill \\ https://doi.org/10.5281/zenodo.1973118 \cite{CarsonCode}.

\subsubsection*{Authors' Contributions}

JC developed and implemented the code, and drafted the manuscript. All authors contributed to designing the research, analysing the results, and writing the manuscript. All authors gave final approval for publication.

\subsubsection*{Competing Interests}

We have no competing interests.

\subsubsection*{Funding}
JC was funded by an EPSRC PhD studentship. RDW was supported by the EPSRC-funded Past Earth Network (Grant number EP/M008363/1). MC is supported by the Belgian National Fund of Scientific Research.

\subsubsection*{Acknowledgements}
The authors thank the referees for their constructive comments and suggestions. JC also thanks Caitlin Buck for discussions during the completion of this work.





\end{document}